\documentclass[a4paper,oneside,12pt]{elsarticle}
\usepackage{amsthm}
\usepackage{amsmath}
\usepackage{graphicx}%
\usepackage{amsfonts}%
\usepackage{amssymb}
 \usepackage{verbatim}
 \usepackage{float}
\usepackage{rotating}
\usepackage{amsmath,amsthm,amssymb}

\newtheorem{defi}{Definition}

\newtheorem{prop}{Proposition}
\newtheorem{rem}{Remark}

\begin{document}

\title{\huge{\textbf{Regime Switching Volatility Calibration by the Baum-Welch Method}}\\
\large{\textbf{by\\
Sovan Mitra}}} \maketitle
\section*{\large{Abstract}}
Regime switching volatility models provide a tractable method of
modelling stochastic volatility. Currently the most popular method
of regime switching calibration is the Hamilton filter. We propose
using the Baum-Welch algorithm, an established technique from
Engineering, to calibrate regime switching models instead. We
demonstrate the Baum-Welch algorithm and discuss the significant
advantages that it provides compared to the Hamilton filter. We
provide computational results of calibrating the Baum-Welch filter
to S\&P 500 data and validate its performance in and out of
sample.
\\\\
\textbf{Key words}: Regime switching, stochastic volatility,
calibration, Hamilton filter, Baum-Welch.

\line(1,0){400}

\section{Introduction and Outline}
Regime switching (also known as hidden Markov models (HMM))
volatility models provide a tractable method of modelling
stochastic volatility. Currently the most popular method of regime
switching calibration is the Hamilton filter. However, regime
switching calibration has been tackled in engineering
(particularly for speech processing) for some time using the
Baum-Welch algorithm (BW), where it is the most popular and
standard method of HMM calibration. A review of the Baum-Welch
algorithm can be found in
\cite{levinson2005mms},\cite{juang1991hmm}. The BW algorithm is
increasingly being applied beyond engineering applications (for
instance in bioinformatics \cite{boufounos2004buh}) but has been
hardly applied to financial modelling, especially to regime
switching stochastic volatility models.

Unlike the Hamilton filter, the BW algorithm is capable of
determining the entire set of HMM parameters from a sequence of
observation data. Furthermore, BW is a complete estimation method
since it also provides the required optimisation method to
determine the parameters by MLE.

The outline of the paper is as follows. Firstly, we introduce
regime switching volatility models and the Hamilton filter. In the
next section we introduce the Baum-Welch method, describing the
algorithm even for multivariate Gaussian mixture observations. We
then conduct a numerical experiment to verify the Baum-Welch
method's to detect regimes for the S\&P 500 index. We finally end
with a conclusion.

\section{Regime Switching Volatility Model and Calibration}\label{Regime Switching Volatility Model Introduction}

\subsection{Regime Switching Volatility}
Wiener process driven stochastic volatility models capture price
and volatility dynamics more successfully compared to previous
volatility models. Specifically, such models successfully capture
the short term volatility dynamics. However, for longer term
dynamics and fundamental economic changes (e.g. ``credit crunch"),
no mechanism existed to address the change in volatility dynamics
and it has been empirically shown that volatility is related to
long term and fundamental conditions. Bekaert in
\cite{bekaert2006gva} claims that volatility
changes are caused by economic reforms, for 
example on Black Wednesday the pound sterling was withdrawn from
the ERM (European Exchange Rate Mechanism), causing a sudden
change in value of the pound sterling \cite{brooks2002tns}.
Schwert
\cite{schwert1989dsv} empirically shows that volatility 
increases during financial crises.

A class of models that address fundamental and long term
volatility modelling is the regime switching model (or hidden
Markov model) e.g. as discusssed in
\cite{timmermann2000mms},\cite{elliott1997ahm}. In fact, Schwert
suggests in \cite{schwert1989dsv} that volatility changes during
the Great Depression can be accounted
for by a regime change such as in Hamilton's regime switching model \cite{hamilton1989nae}.  
Regime switching is considered a tractable method of modelling
price dynamics and does not violate Fama's ``Efficient Market
Hypothesis"\cite{fama1965bsm}, which claims that price processes
must follow a Markov process. Hamilton \cite{hamilton1989nae} was
the first to introduce regime switching models, which was applied
to specifically model fundamental economic changes.

For regime switching models, generally the return distribution
rather than  the continuous time process is specified. A typical
example of a regime switching model is Hardy's model
\cite{hardy2001rsm}:
\begin{eqnarray}
log((X(t+1)/X(t))|i) \sim \mathcal{N}(u_{i},\varphi_{i}), i \in
\{1,..,R \}, \label{Hardy eqn}
\end{eqnarray}
where
\begin{itemize}
  \item $\varphi_{i}$ and $u_{i}$ are constant for the duration of the
regime;
    \item i denotes the current regime (also called the Markov
  state or hidden Markov state);
\item R denotes the total number of regimes;
\item a transition matrix \textbf{A} is specified.
\end{itemize}
For Hardy's model the regime changes discretely in monthly time
steps but stochastically, according to a Markov process.

Due to the ability of regime switching models to capture long term
and fundamental changes, regime switching models are primarily
focussed on modelling the long term behaviour, rather than the
continuous time dynamics. Therefore regime switching models switch
regimes over time periods of months, rather than switching in
continuous time. Examples of regime switching models that model
dynamics over shorter time periods are Valls-Pereira et
al.\cite{vallspereira2004pva}, who propose a regime switching
GARCH process, while Hamilton and Susmel \cite{hamilton1994ach}
give a regime switching ARCH process. Note that economic variables
other than stock returns, such as inflation, can also be modelled
using regime switching models.

Regime switching has been developed by various researchers. For
example, Kim and Yoo \cite{kim1995nic} develop a multivariate
regime switching model for coincident economic indicators. Honda
\cite{honda2003opc} determines the optimal portfolio choice in
terms of utility for assets following GBM but with  continuous
time regime switching
mean returns. 
Alexander and Kaeck \cite{alexander2007rdd} apply regime switching
to credit default swap spreads, Durland and McCurdy
\cite{durland1994ddt} propose a model with a transition matrix
that specifies state durations.

The theory of Markov models (MM) and Hidden Markov models (HMM)
are methods of mathematically modelling time varying dynamics of
certain statistical processes, requiring a weak set of assumptions
yet allow us to deduce a significant number of properties. MM and
HMM model a stochastic process (or any system) as a set of states
with each state possessing a set of signals or observations. The
models have been used in diverse applications such as  economics
\cite{sola2002tvs}, queuing theory \cite{salih2006maa},
engineering \cite{trentin2001sha} and biological modelling
\cite{melodelima2006cpi}. Following Taylor \cite{taylor1984ism} we
define a Markov model:
\begin{defi}
A Markov model is a stochastic process $X(t)$ with a countable set
of states and possesses the Markov property:
\begin{eqnarray}
p(q_{t+1}=j \mid q_{1},q_{2},..,q_{t}=i)=p(q_{t+1}=j \mid
q_{t}=i),
\end{eqnarray}
where
\begin{itemize}
  \item $q_{t}$ is the Markov state (or regime) at time t of
  $X(t)$;
  \item i and j are specific Markov states.
\end{itemize}
\end{defi}
As time passes the process may remain or change to another state
(known as state transition). The state transition probability
matrix (also known as the \emph{transition kernel or stochastic
matrix}) $\mathbf{A}$, with elements $a_{ij}$, tells us the
probability of the process changing to state \emph{j} given that
we are now in state \emph{i}, that is $a_{ij}=p(q_{t+1}=j \mid
q_{t}=i)$. Note that $a_{ij}$ is subject to the standard
probability constraints:
\begin{eqnarray}
0 \leq a_{ij} & \leq &  1, \forall  i,j,\\
\sum^{\infty}_{j=1} a_{ij} &=& 1, \forall  i.
\end{eqnarray}
We assume that all probabilities are stationary in time. From the
definition of a MM the following proposition follows:
\begin{prop} 
A  Markov model is completely defined once the following
parameters are known:
\begin{itemize}
  \item \emph{R}, the  total number of regimes or (hidden) states; 
\item state transition probability matrix $\mathbf{A}$ of size $R\times R$. Each element is
$a_{ij}=p(q_{t+1}=j|q_{t}=i)$, where i refers to the matrix row
number and j to the column number of $\mathbf{A}$;
  \item initial (t=1) state probabilities $\pi_{i}=p(q_{1}=i), \forall i$.
  \end{itemize}
\end{prop}
A hidden Markov model is simply a Markov model where we assume (as
a modeller) we do not observe the Markov states. Instead of
observing the Markov states (as in standard Markov models) we
detect observations or time series data where each observation is
assumed to be a function of the hidden Markov state, thus enabling
statistical inferences about the HMM.
Note that in a HMM it is the states which must be governed by a
Markov process, not the observations and throughout the thesis we
will assume one observation occurs after one state
transition. 

\begin{prop}\label{HMM def prop}
A hidden Markov model is fully defined  when
the parameter set $\{\mathbf{A},\mathbf{B},\mathbf{\pi}\}$ are known: 
\begin{itemize}
  \item \emph{R}, the  total number of (hidden) states or regimes;
\item \textbf{A}, the (hidden) state transition matrix of size $R\times R$.
Each element is $a_{ij}=p(q_{t+1}=j|q_{t}=i)$;
  \item initial (t=1) state probabilities $\pi_{i}=p(q_{1}=i), \forall i$;
\item $\textbf{B}$, the observation matrix, where each entry is
$b_{j}(O_{t})=p(O_{t}|j)$ for observation $O_{t}$. For
$b_{j}(O_{t})$ is typically defined to follow some continuous
distribution e.g. $b_{j}(O_{t}) \sim
\mathcal{N}(u_{j},\varphi_{j})$.
\end{itemize}
\end{prop}

\subsection{Current Calibration Method: Hamilton Filter}
In financial mathematics or economic literature the standard
calibration method for regime switching models is the Hamilton
filter \cite{hamilton1989nae}, which works by maximum likelihood
estimation (MLE). MLE is a method of estimating a set of
parameters of a statistical model ($\Theta$) given some time
series or empirical observations $O_{1},O_{2},...,O_{T}$. MLE
determines $\Theta$ by firstly determining the likelihood function
$\mathcal{L}(\Theta)$, then maximising $\mathcal{L}(\Theta)$ by
varying $\Theta$ through a search or an optimisation method.

A statistical model with known parameter values can determine the
probability of an observation sequence $O=O_{1}O_{2}...O_{T}$. MLE
does the opposite;  we numerically maximise the parameter values
of our model $\Theta$ such that we maximise the probability of the
observation sequence $O=O_{1}O_{2}...O_{T}$. To achieve this the
MLE method makes two assumptions:
\begin{enumerate}
\item In maximising $\mathcal{L}(\Theta)$ the local optimum is also the global optimum
 (although this is generally not true in reality). The optimal values for $\Theta$ are in a search space
 of the same dimensions as $\Theta$. Hamilton in \cite{hamilton1994tsa} 
gives a survey of various MLE maximisation techniques such as the
Newton-Raphson method;

\item The observations $O_{1},O_{2},...,O_{T}$ are statistically independent. Note that for
Markov models we assume the conditional observations\\
$(O_{t}|O_{t-1}),(O_{t-1}|O_{t-2}),(O_{t-2}|O_{t-3})...$ are
independent.

\end{enumerate}

For a regime switching process the general likelihood function
$\mathcal{L}(\Theta)$ is:
\begin{eqnarray*}
\mathcal{L}(\Theta) & = &
f(O_{1}|\Theta)f(O_{2}|\Theta,O_{1})f(O_{3}|\Theta,O_{1},O_{2})\\
& \cdots & f(O_{T}|\Theta,O_{1},O_{2},...,O_{T-1}),
\end{eqnarray*}
where $f(O_{(.)}|\Theta)$ is the probability of $O_{(.)}$, given
model parameters $\Theta$. Now by properties of logarithms we
have:
\begin{eqnarray}
log(\mathcal{L}(\Theta)) & = &
log(f(O_{1}|\Theta))+log(f(O_{2}|\Theta,O_{1}))+\cdots\\\label{RS
gen MLE eqn}
&+& log(f(O_{T}|\Theta,O_{1},O_{2},...,O_{T-1})). 
\end{eqnarray}
Hamilton proposes a likelihood function for regime switching models, the Hamilton filter. 
 As an example, if we assume we have a two regime model
with each regime having a lognormal return distribution, we wish
to determine parameters
$\Theta=\{u_{1},u_{2},\varphi_{1},\varphi_{2},a_{12},a_{21}\}$.
Note that in this simple HMM $a_{22}=1-a_{12}$ and
$a_{11}=1-a_{21}$ therefore we do not need to estimate
$a_{22},a_{11}$ in $\Theta$.

To obtain $f(O_{t}|\Theta)$ in equation (\ref{RS gen MLE eqn}) for
$t>1$, Hamilton showed it could be calculated by a recursive
filter. We observe
the relation: 
\begin{eqnarray}
f(O_{t}|\Theta,O_{1},O_{2},...,O_{t-1}) =
\sum_{q_{t}=1}^{2}\sum_{q_{t-1}=1}^{2}
f(q_{t},q_{t-1},O_{t}|\Theta,O_{1},...,O_{t-1}).
\end{eqnarray}
Now using the relation:\footnote{Note: following discussions with
Prof.Rabiner \cite{RabinerPrivCom} on the equation for
$p(O,Q|\Theta)$ it was concluded
that the equation for $p(O,Q|\Theta)$ in Rabiner's paper \cite{rabiner1989thm} is incorrect.} 
\begin{eqnarray}
p(O,Q|\Theta)=p(O|\Theta,Q)p(Q|\Theta),
\end{eqnarray}
where $Q=q_{1}q_{2}...$ represents some arbitrary state sequence,
we make the substitution
$f(q_{t},q_{t-1},O_{t}|\Theta,O_{1},...,O_{t-1})$
\begin{eqnarray}\label{Ham filt sub eqn}
= p(q_{t-1}|\Theta,O_{1},...,O_{t-1}) \times
p(q_{t}|q_{t-1},\Theta)
 \times  f(O_{t}|q_{t},\Theta).
\end{eqnarray}
Therefore\\
$f(O_{t}|\Theta,O_{1},O_{2},...,O_{t-1})$
\begin{eqnarray}
= \sum_{q_{t}=1}^{2}\sum_{q_{t-1}=1}^{2}
p(q_{t-1}|\Theta,O_{1},...,O_{t-1})\times
p(q_{t}|q_{t-1},\Theta)\times f(O_{t}|q_{t},\Theta),
\label{Hamilton Filter}
\end{eqnarray}
where
\begin{itemize}
  \item $p(q_{t}|q_{t-1},\Theta)=p(q_{t}=j|q_{t-1}=i,\Theta)$ represents the transition
  probability $a_{ij}$ we wish to estimate;
  \item
 $f(O_{t}|q_{t}=i,\Theta)=p_{i}(O_{t})$
where $p_{i}(\cdot) \sim \mathcal{N}(u_{i},\varphi_{i})$ the
Gaussian probability density  function for state $i$, whose
parameters $u_{i},\varphi_{i}$  we wish to
  estimate.
\end{itemize}
The parameters
$\Theta=\{u_{1},u_{2},\varphi_{1},\varphi_{2},a_{12},a_{21}\}$ are
obtained by maximising the likelihood function using some chosen
search method.

To calculate $f(O_{t}|\Theta,O_{1},O_{2},...,O_{t-1})$ we require
the probability \\ $p(q_{t-1}|\Theta,O_{1},O_{2},...,O_{t-1})$ in
equation (\ref{Hamilton Filter}) (summed over  two different
values of $q_{t-1}$  in the summations in equation (\ref{Hamilton
Filter})). This can be achieved through recursion, that is the
probability $p(q_{t-1}|\Theta,O_{1},..,O_{t-1})$ can be obtained
from $p(q_{t-2}|\Theta,O_{1},..,O_{t-2})$:
\begin{eqnarray}\label{recursion eqn}
p(q_{t-1}|\Theta,O_{1},O_{2},...,O_{t-1})=\frac{\left(\sum_{i=1}^{2}
f(q_{t-1},q_{t-2}=i,O_{t-1}|\Theta,O_{1},...,O_{t-2})\right)}
{f(O_{t-1}|\Theta,O_{1},...,O_{t-2})}.
\end{eqnarray}
The denominator of equation (\ref{recursion eqn}) is obtained from
the
previous period of\\
 $f(O_{t}|\Theta,O_{1},O_{2},...,O_{t-1})$ (in
other words $f(O_{t-1}|\Theta,O_{1},O_{2},...,O_{t-2})$) so by
inspecting equation (\ref{Hamilton Filter}) we can see it is a
function of $p(q_{t-2}|\Theta,O_{1},...,O_{t-2})$. The numerator
of equation (\ref{recursion eqn}) is obtained from calculating
equation (\ref{Ham filt sub eqn}) for the previous time period,
which is also a function of $p(q_{t-2}|\Theta,O_{1},...,O_{t-2})$.

To start the recursion of equation (\ref{recursion eqn}) at
$p(q_{1}=i|O_{1},\Theta)$ we require $f(O_{1}|\Theta)$. Hamilton
assumes the Markov chain has been running sufficiently long enough
so that we can make the following assumption about our
observations $O_{1},O_{2},...,O_{T}$. Technically, Hamilton
assumes the observations $O_{1},O_{2},...,O_{T}$ are all drawn
from the Markov chain's invariant distribution. If a Markov chain
has been running for a sufficiently long time, the following
property of Markov chains can be applied:
\begin{eqnarray}
\eta_{j} &=& \lim_{t\rightarrow \infty}p(q_{t}=j|q_{1}=i),\qquad
\forall i,j=1,2,..,R,
\label{Hamilton inf state}\\
\mbox{where } \sum_{j=1}^{R}\eta_{j} &=& 1,\eta_{j}>0.
\end{eqnarray}
The probability $\eta_{j}$ tells us in the long run $(t\rightarrow
\infty)$ the (unconditional) probability of being in state j and
this probability is independent of the initial state (at time
t=1). An important interpretation of $\eta_{j}$ is as the fraction
of time spent in state j in the long run. 
 Therefore the
probability of state j is simply $\eta_{j}$ and so:
\begin{eqnarray}
f(O_{1}|\Theta) &=& f(q_{1}=1,O_{1}|\Theta)+f(q_{1}=2,O_{1}|\Theta),\\
\mbox{where }f(q_{1}=i,O_{1}|\Theta) &=& \eta_{i}p_{i}(O_{1}).
\end{eqnarray}
We can therefore calculate $p(q_{1}=i|O_{1},\Theta)$:
\begin{eqnarray}
p(q_{1}=i|O_{1},\Theta) &=& \frac{f(q_{1}=i,O_{1}|\Theta)}{f(O_{1}|\Theta)},\\
&=&
\frac{\eta_{i}p_{i}(O_{1})}{\eta_{1}p_{1}(O_{1})+\eta_{2}p_{2}(O_{1})}.
\end{eqnarray}
Furthermore it can be proved for a two state HMM that:
\begin{eqnarray*}
\eta_{1} &=& a_{21}/(a_{12}+a_{21}),\\
\eta_{2} &=& 1-\eta_{1}.
\end{eqnarray*}
Therefore $p(q_{1}=i|O_{1},\Theta)$ can be obtained from
estimating the parameter set
$\Theta=\{u_{1},u_{2},\varphi_{1},\varphi_{2},a_{12},a_{21}\}$,
which is obtained by a chosen search method.

The advantages of Hamilton's filter method are firstly we do not
need to specify or determine the initial probabilities, therefore
there are fewer parameters to estimate (compared to the
alternative Baum-Welch method). Therefore the MLE parameter
optimisation will be over a lower dimension search space.
Secondly, the MLE equation is simpler to understand and so easier
to implement compared to other calibration methods.

\section{Baum-Welch Algorithm}\label{Baum-Welch Algorithm section}
The Baum-Welch (BW) is a complete estimation method since it also
provides the required optimisation method to determine the
parameters by MLE. We will now explain the BW algorithm and to do
so we must first explain the forward algorithm, which we will do
now.

\subsection{Forward Algorithm}
The forward algorithm calculates $p(O|M)$, the probability of a
fixed or observed\\ sequence
 O=$O_{1}O_{2}...O_{T}$, 
given all the HMM parameters denoted by
$M=\{\mathbf{A},\mathbf{B},\mathbf{\pi} \}$. We recall from the
definition of HMM that the probability of each observation
$p(O_{t})$ will change depending on the state at time t ($q_{t}$).
Hence the most straightforward way to calculate $p(O|M)$ is:
\begin{eqnarray}
p(O|M) &=& \sum_{\mbox{all Q}} p(O,Q|M),\\
&=& \sum_{\mbox{all Q}}p(O|M,Q).p(Q|M),\\
&=&
\sum_{\mbox{all Q}}
\pi_{q_{1}}b_{q_{1}}(O_{1}).a_{q_{1}q_{2}}b_{q_{2}}(O_{2})....a_{q_{T-1}q_{T}}
b_{q_{T}}(O_{T}), \label{p(O|M)}\\
\mbox{where } 
p(O|M,Q)&=&
b_{q_{1}}(O_{1}).b_{q_{2}}(O_{2}).....b_{q_{T}}(O_{T}).
\end{eqnarray}
Here ``all Q" means all possible state sequences
$q_{1}q_{2}...q_{T}$ that could account for observation sequence
O, $b_{(.)}(O_{(.)})$ is defined in proposition \ref{HMM def
prop}, $p(O|M,Q)$ is the probability of the observed sequence O,
given it is along one single state sequence
$Q=q_{1}q_{2}...q_{T}$ and for HMM M. We must sum equation (\ref{p(O|M)}) 
over all possible Q state sequences, requiring $R^{T}$
computations and so this is computationally infeasible even for
small R and T.

To overcome the computational difficulty of calculating $p(O|M)$
in equation (\ref{p(O|M)}) we apply the forward algorithm, which
uses recursion (dynamic
programming). 
The forward algorithm only requires computations of the order
$R^{2}T$ and so is significantly faster than calculating equation
(\ref{p(O|M)}) for large R and T.

Let us define the forward variable $\kappa_{t}(i)$:
\begin{eqnarray}
\kappa_{t}(i)=p(O_{1}O_{2}...O_{t},q_{t}=i|M).
\end{eqnarray}
Given the HMM M, $\kappa_{t}(i)$ is the probability of the joint
observation upto time t of
 $O_{1}O_{2}...O_{t}$ and the state at time t is $i$
 i.e. $q_{t}=i$. 
If we can determine $\kappa_{T}(i)$ we can calculate $p(O|M)$
since:
\begin{eqnarray}
p(O|M)=\sum_{i=1}^{R}\kappa_{T}(i).
\end{eqnarray}

Now $\kappa_{t+1}(j)$ can be expressed in terms of
$\kappa_{t}(i)$, therefore we can calculate $\kappa_{t+1}(j)$ by
recursion:
\begin{eqnarray}\label{forward alg eqn}
\kappa_{t+1}(j)=\left[\sum_{i=1}^{R}\kappa_{t}(i)a_{ij} \right]b_{j}(O_{t+1}), 1\leq t \leq T-1. 
\end{eqnarray}
The variable $\kappa_{t+1}(j)$ in equation (\ref{forward alg eqn})
can be understood as follows: $\kappa_{t}(j)a_{ij}$ is the
probability of the joint event $O_{1}....O_{t}$ is observed, the
state at time t is $i$ and state $j$ is reached at time t+1. If we
sum this probability over all R possible states for $i$,  we get
the probability of $j$ at t+1 accompanied with all previous
observations from $O_{1}O_{2}...O_{t}$ only. Thus to get
$\kappa_{t+1}(j)$ we must multiply by $b_{j}(O_{t+1})$ so that we
have all observations $O_{1}...O_{t+1}$.

Therefore the recursive algorithm is as follows:  
\begin{enumerate}
\item Initialisation:
\begin{eqnarray}
\kappa_{1}(i)=\pi_{i}b_{i}(O_{1}), 1\leq i \leq R.
\end{eqnarray}
\item Recursion:
\begin{eqnarray}
\kappa_{t+1}(j)=\left[\sum_{i=1}^{R}\kappa_{t}(i)a_{ij}
\right]b_{j}(O_{t+1}), 1\leq t \leq T-1.
\end{eqnarray} 
\item Termination: t+1=T.
\item Final Output:
\begin{eqnarray}
p(O|M)=\sum_{i=1}^{R}\kappa_{T}(i).
\end{eqnarray}
\end{enumerate}
At t=1 no sequence exists but we initialise the recursion with
$\pi_{i}$ to determine $\kappa_{1}(i)$.

\subsection{Baum-Welch Algorithm}
Having explained the forward algorithm we can now explain the BW
algorithm. Using observation sequence O, the BW algorithm
iteratively calculates the HMM parameters
$M=\{\mathbf{A},\mathbf{B},\mathbf{\pi} \}$. Specifically, BW
estimates $M=\{a_{ij},b_{j}(\cdot),\pi_{i}\}  \forall i,j$,
denoted respectively by
$\overline{M}=\{\overline{a}_{ij},\overline{b}_{j}(\cdot),
\overline{\pi}_{i} \}$, such that it maximises the likelihood of
$p(O|\overline{M})$. No method of analytically finding the
globally optimal $\overline{M}$ exists. However it has been
theoretically proven BW is guaranteed to find
the local optimum \cite{rabiner1989thm}. 

Let us define $\psi_{t}(i,j)$:
\begin{eqnarray}
 \psi_{t}(i,j)=p(q_{t}=i,q_{t+1}=j\mid O,M).
\end{eqnarray}
The variable $\psi_{t}(i,j)$ is the probability of being in state
i at time t and state j at time t+1, given the HMM parameters M
and the observed observation sequence O. We
can re-express $\psi_{t}(i,j)$ as:
\begin{eqnarray}\label{psi eqn3}
\psi_{t}(i,j) &=& p(q_{t}=i,q_{t+1}=j\mid O,M),\\ \label{psi eqn2}
             &=& \frac{p(q_{t}=i,q_{t+1}=j,O \mid M)}{p(O|M)}.
\end{eqnarray}
Now we can re-express equation (\ref{psi eqn2}) using the forward
variable\\ $\kappa_{t}(i)=p(O_{1}O_{2}...O_{t},q_{t}=i|M)$ and
using analogously the so called backward variable $\varrho_{t+1}(i)$:
\begin{eqnarray}
\varrho_{t}(i)=p(O_{t+1}O_{t+2}...O_{T}|q_{t}=i,M),\\
\mbox{so that }
\varrho_{t+1}(i)=p(O_{t+2}O_{t+3}...O_{T}|q_{t+1}=i,M).
\end{eqnarray}
The backward variable $\varrho_{t}(i)$ is the probability of the
partial observed observation sequence from time t+1 to the end T,
given M and the state at time t is i. It is calculated in a
similar recursive method to the forward variable using  the
backward algorithm (see
\cite{rabiner1989thm} for more details). 
 Hence we can rewrite
$\psi_{t}(i,j)$ as
\begin{eqnarray}\label{psi eqn}
\psi_{t}(i,j) &=&
\frac{\kappa_{t}(i)a_{ij}b_{j}(O_{t+1})\varrho_{t+1}(j)}{p(O|M)}.
\end{eqnarray}
We can also rewrite the denominator $p(O|M)$ in terms of the
forward and backward variables, so that $\psi_{t}(i,j)$ is
entirely expressed in terms of
$\kappa_{t}(i),a_{ij},b_{j}(O_{t+1}),\varrho_{t+1}(j)$:
\begin{eqnarray}
p(O|M) = \sum_{i=1}^{R}\sum_{j=1}^{R}
\kappa_{t}(i)a_{ij}b_{j}(O_{t+1})\varrho_{t+1}(j).
\end{eqnarray}

Now let us define $\Gamma_{t}(i)$:
\begin{eqnarray}
\Gamma_{t}(i)  & = & p(q_{t}=i|O,M),\\ \label{Gamma t}
               & = & \sum_{j=1}^{R}\psi_{t}(i,j).
\end{eqnarray}
Equation (\ref{Gamma t}) can be understood from the definition of
$\psi_{t}(i,j)$ in equation (\ref{psi eqn3});  summing
$\psi_{t}(i,j)$ over all j must give $p(q_{t}=i|O,M)$, the
probability in state i at time t, given the observation sequence O
and model M. Now if we sum $\Gamma_{t}(i)$ from t=1 to T-1 it
gives us $\Upsilon(i)$, the expected number of transitions made
from state i:
\begin{eqnarray}
\Upsilon(i)=\sum^{T-1}_{t=1} \Gamma_{t}(i).
\end{eqnarray}
If we sum $\Gamma_{t}(i)$ from t=1 to T it gives us
$\vartheta(i)$, the expected number of times state i is visited:
\begin{eqnarray}
\vartheta(i)=\sum^{T}_{t=1} \Gamma_{t}(i).
\end{eqnarray}

We are now in a position to estimate $\overline{M}$. The variable
$\overline{a}_{ij}$ is estimated as the expected number of
transitions from state i to state j divided by the expected number
of transitions from state i:
\begin{eqnarray}\label{aij est eqn}
\overline{a}_{ij}= \dfrac{\sum_{t=1}^{T-1} \psi_{t}(i,j)}{\Upsilon(i)}.
\end{eqnarray}
The variable $\overline{\pi}_{i}$ is estimated as the expected number of times in state i at time t=1: 
\begin{eqnarray}\label{pi est eqn}
\overline{\pi}_{i}=\Gamma_{1}(i).
\end{eqnarray}
The variable $\overline{b}_{j}(\tilde{s})$ is estimated as the
expected number of times in state j and observing a particular
signal $\tilde{s}$, divided by the expected number of times in
state j:
\begin{eqnarray}\label{bj est eqn}
\overline{b}_{j}(\tilde{s}) = \dfrac{\displaystyle\sum_{t=1}^{T}
\Gamma_{t}(j)'}{\vartheta(j)},
\end{eqnarray} 
where  $\Gamma_{t}(j)'$ is $\Gamma_{t}(j)$ with condition
$O_{t}=\tilde{s}$.

We can now describe our BW algorithm:
\begin{enumerate}
\item Initialisation:\\
Input initial values of $\overline{M}$ (otherwise randomly
initialise) and
calculate $p(O|\overline{M})$ using the forward algorithm. 
\item Estimate new values of $\overline{M}$:\\
Iterate until convergence:
\begin{enumerate}
\item Using current $\overline{M}$ calculate variables
$\overline{\kappa}_{t}(i),\overline{\varrho}_{t+1}(j)$ by the
forward and backward algorithm and then calculate
$\overline{\psi}_{t}(i,j)$ as in equation (\ref{psi eqn}).
\item Using calculated $\overline{\psi}_{t}(i,j)$ in (a) determine new estimates of
$\overline{M}$ using equations (\ref{Gamma t})-(\ref{bj est eqn}).
\item Calculate $p(O|\overline{M})$ with new $\overline{M}$ values using the
forward algorithm. 
\item Goto step 3 if two consecutive calculations of
$p(O|\overline{M})$ are equal (or converge within a specified
range). Otherwise repeat iterations: goto (a).
\end{enumerate}
\item Output $\overline{M}$.
\end{enumerate}
The BW algorithm is started with initial estimates of
$\overline{M}=(\mathbf{\overline{A}},\mathbf{\overline{B}},\overline{\pi})$.
These estimates in turn are used to calculate the right hand side
of equations (\ref{aij est eqn}),(\ref{pi est eqn}) and (\ref{bj
est eqn}) to give the next new estimate of
$\overline{M}=(\mathbf{\overline{A}},\mathbf{\overline{B}},\overline{\pi})$.
We consider the new estimate $\overline{M_{n}}$ to be a better
estimate than the previous estimate $\overline{M_{p}}$, if
$p(O|\overline{M_{n}})>p(O|\overline{M_{p}})$, with both
probabilities calculated via the forward algorithm. In other
words, we prefer the $\overline{M}$ that increases the probability
of observation O occurring.

If $p(O|\overline{M_{n}})>p(O|\overline{M_{p}})$ then the
iterative calculation is repeated with $\overline{M_{n}}$ as the
input. Note that at the end of step two, if the algorithm
re-iterates then inputting the new $\overline{M}$ at step 2a means
we will get a new set of $\overline{M}$ after executing 2b. The
iteration is stopped when
$p(O|\overline{M_{n}})=p(O|\overline{M_{p}})$ or is arbitrarily
close enough and at this point the BW algorithm finishes.

Since the BW algorithm has been proven to always converge to the
local optimum, the BW will output the local optimum. We also note
that correct choice of R is important since $p(O|M)$ changes as M
changes for a fixed O, however this disadvantage is common to all
MLE methods.

\subsection{Multivariate Gaussian Mixture Baum-Welch Calibration}
To account for the variety of empirical distributions possible for
various assets and capturing asymmetric properties arising from
volatility (such as fat tails), we model each regime's
distribution by a two component multivariate Gaussian mixture
(GM), which is a mixture of two multinormal distributions.
\begin{defi}
(Multinormal Distribution) Let $\mathbf{X}=(X_{1},X_{2}...,X_{n})$
be an n-dimensional random vector where each dimension is a random
variable. 
Let {\boldmath $u$}=$(u_{1},u_{2}...,u_{n})$ represent an n
dimensional vector of means, $\mathbf{\Sigma}$ represent an $n
\times n$ covariance matrix. We say $\mathbf{X}$ follows a
multinormal distribution if
\begin{eqnarray}
\mathbf{X \sim \mathcal{N}_{n}(\mbox{\boldmath$u$},\Sigma)},  
\end{eqnarray}
which may be alternatively written as
\begin{eqnarray}
\begin{pmatrix}
  X_{1} \\
  X_{...} \\
  X_{n}
\end{pmatrix}
\sim \mathcal{N}_{n} \left(\begin{pmatrix}
  u_{1} \\
  u_{...} \\
  u_{n}
\end{pmatrix},\begin{pmatrix}
  \varphi_{11} & \varphi_{...} & \varphi_{1n} \\
  \varphi_{...} & \varphi_{...} & \varphi_{...} \\
  \varphi_{n1} & \varphi_{...} & \varphi_{nn}
\end{pmatrix} \right).
\end{eqnarray}
The probability of $\mathbf{X}$ is
\begin{eqnarray}
p(\mathbf{X})=\dfrac{1}{2\pi
\sqrt{det(\mathbf{\Sigma})}}exp\left(-\frac{1}{2}(\mathbf{X}-\mbox{\boldmath$u$})^{T}\mathbf{\Sigma}^{-1}(\mathbf{X}-\mbox{\boldmath$u$})\right),
\end{eqnarray} 
where $det(\mathbf{\Sigma})$ denotes the determinant of
$\mathbf{\Sigma}$.
\end{defi}
\begin{defi}
(Multivariate Gaussian Mixture) 
A multivariate Gaussian mixture consists of a mixture of K
multinormal distributions, spanning n-dimensions. It is defined
by:
\begin{eqnarray}
\mathbf{X}  \sim  c_{1}
\mathbf{\mathcal{N}_{n}}(\mathbf{u}_{1},\mathbf{\Sigma}_{1})+...+c_{K}\mathbf{\mathcal{N}_{n}}(\mathbf{u}_{K},\mathbf{\Sigma}_{K}),
\end{eqnarray}
where $c_{k}$ are weights and
\begin{eqnarray}
  \sum_{k=1}^{k=K}c_{k} = 1, c_{k}\geq 0.
\end{eqnarray}
The term $p_{gmm}(\mathbf{X})$ denotes the probability of a
multivariate Gaussian mixture variable $\mathbf{X}$ and is defined
as
\begin{eqnarray}\label{p_gmm}
p_{gmm}(\mathbf{X})  &=& \sum^{K}_{k=1} c_{k} p_{k}(\mathbf{X}),
\end{eqnarray}
where $p_{k}(\mathbf{X}) \sim
\mathbf{\mathcal{N}_{n}}(\mathbf{u}_{k},\mathbf{\Sigma}_{k})$.
\end{defi}
If we model a stochastic process \textbf{X} by a Gaussian mixture
for each regime then for a given regime j we have:
\begin{eqnarray}
\mathbf{X} & \sim &
c_{j1}\mathbf{\mathcal{N}_{n}}(\mathbf{u}_{j1},\mathbf{\Sigma}_{j1})+...+c_{jK}\mathbf{\mathcal{N}_{n}}(\mathbf{u}_{jK},\mathbf{\Sigma}_{jK}).
\end{eqnarray}
The probability of $\mathbf{X}$ for a given regime j,
$p_{gmm}(\mathbf{X})_{j}$, is:
\begin{eqnarray}
p_{gmm}(\mathbf{X})_{j} & = & \sum^{k=K}_{k=1}
c_{jk}p_{jk}(\mathbf{X}).
\end{eqnarray}
where
\begin{itemize}
\item $p_{jk}(\mathbf{X}) \sim
\mathbf{\mathcal{N}_{n}}(\mathbf{u}_{jk},\mathbf{\Sigma}_{jk}); $

\item $c_{jk}$ are weights for each regime j and
\begin{eqnarray}
\sum_{k=1}^{k=K}c_{jk} = 1,\forall j.
\end{eqnarray}

\end{itemize}
Note that the dimensions of multivariate distribution n are
independent of the number of mixture components K. 

For an n-asset portfolio
$\mathbf{X(t)}=(X_{1}(t),X_{2}(t),...,X_{n}(t))$, where $X_{i}(t)$
represents the stock price of asset i, with each asset following a
Gaussian mixture, the portfolio returns would be modelled by:
\begin{eqnarray}
d\mathbf{X/X} \sim c_{j1}
\mathbf{\mathcal{N}_{n}}(\mathbf{u}_{j1},\mathbf{\Sigma}_{j1}) +
c_{j2}\mathbf{\mathcal{N}_{n}}(\mathbf{u}_{j2},\mathbf{\Sigma}_{j2}).
\end{eqnarray}
For practial calibration purposes we set the multivariate
observation \textit{vector} $\mathbf{O_{t}}$ to annual log
returns:
\begin{eqnarray}
\mathbf{O_{t}}=log(\mathbf{X}(t+\Delta t)/\mathbf{X}(t)),
\end{eqnarray}
 where
$\Delta t$=1 year.

Combining GM with HMM gives us a GM-HMM (Gaussian mixture HMM)
model and the BW algorithm can be adapted to it: Gaussian mixture BW (GM-BW). 
For $\textbf{O}_{t}$ our observation (vector) at time t we model
$b_{j}(\textbf{O})$ by GM:
\begin{eqnarray}
b_{j}(\textbf{O})=p_{gmm}(\textbf{O})_{j}. 
\end{eqnarray}
The BW algorithm for calculating $\mathbf{A},\pi_{i}$ remains the
same;  for \textbf{B} we have a GM. We would like to obtain the GM
mixture coeffficents $c_{jk}$, mean vectors
$\mbox{\textbf{\textit{u}}}_{jk}$ and covariance matrices
$\mathbf{\Sigma}_{jk}$ whose estimates are $\overline{c}_{jk}$,
$\overline{\mbox{\textbf{\textit{u}}}}_{jk}$ and
$\overline{\mathbf{\Sigma}}_{jk}$ respectively. These can be
incorporated within the BW algorithm as detailed by Rabiner
\cite{rabiner1989thm}:
\begin{eqnarray}\label{GM ujk eqn}
\overline{\mbox{\textbf{\textit{u}}}}_{jk} &=& \dfrac{\sum_{t=1}^{T}\Gamma_{t}(j,k).\textbf{O}_{t}}{\sum_{t=1}^{T}\Gamma_{t}(j,k)},\\
\overline{\mathbf{\Sigma}}_{jk} &=&
\dfrac{\sum_{t=1}^{T}\Gamma_{t}(j,k).(\textbf{O}_{t}-\overline{\mbox{\textbf{\textit{u}}}}_{jk}
)(\textbf{O}_{t}-\overline{\mbox{\textbf{\textit{u}}}}_{jk})^{T}}{\sum_{t=1}^{T}\Gamma_{t}(j,k)},\\\label{GM
cjk eqn} \overline{c}_{jk} &=&
\dfrac{\sum_{t=1}^{T}\Gamma_{t}(j,k)}{\sum_{t=1}^{T}
\sum_{k=1}^{K} \Gamma_{t}(j,k)},\\
\mbox{where }\Gamma_{t}(j,k)
& = &
\left[\dfrac{\kappa_{t}(j)\varrho_{t}(j)}{\sum_{j=1}^{N}\kappa_{t}(j)\varrho_{t}(j)}
\right]\left[ \dfrac{
c_{jk}p_{jk}(\textbf{O}_{t})}{p_{gmm}(\textbf{O}_{t})_{j}}
\right].
\end{eqnarray}
 Here $\Gamma_{t}(j,k)$ is the probability at time t of being
in state j with the k mixture component accounting for
$\textbf{O}_{t}$. Using the same logic as in section
\ref{Baum-Welch Algorithm section} (for non mixture distributions)
we can understand equations (\ref{GM ujk eqn})-(\ref{GM cjk eqn}),
for example $\overline{c}_{jk}$ is the expected number of times
the HMM k-th component is in state j divided by the expected
number of times in state j.

It is worth noting that a well known problem in maximum likelihood
estimation of GM is that observations with low variances give
extremely high likelihoods, in which case the likelihood function
does not converge \cite{messina2007hmm}. To overcome this problem
in the univariate case Messina and Toscani \cite{messina2007hmm}
implement Ridolfi's and Idier's \cite{ridolfi2002pml} penalised
maximum likelihood function, which limits the likelihood value of
observations. This is beneficial in \cite{messina2007hmm} because
the observation time scales are of the order of days and therefore
the variance of samples may approach zero. For our applications we
calibrate the GM-HMM to annual return data, therefore the samples
are unlikely to approach variances anywhere near zero.

\subsection{Advantages of Baum-Welch Calibration} 

The BW algorithm has significant advantages over the Hamilton
filter. Firstly, the Hamilton filter requires observation data to
be taken from the invariant distribution in order to estimate the
parameters (see equation (\ref{Hamilton inf state})). To obtain
observations from the invariant distribution implies the number of
state transitions approaches a large limit, so is not suited to
Markov chains that have run for a short time.
Furthermore, the time to reach the invariant distribution
increases with the number of regimes R and the number of Gaussian
mixtures K.

Psaradakis and Sola \cite{psaradakis1996fsp} investigated the
finite sample
 properties of the Hamilton filter for financial data. They
 concluded that samples of at least 400 observations are required
for a simple two state regime switching model where each state's
observation is modelled by a normal distribution.

Secondly, the Hamilton filter has no method of estimating the
initial state probabilities whereas the BW is able to take account
of and estimate initial state probabilities. This has a number of
important consequences:
\begin{enumerate}
\item  BW does not require observations from the invariant
distribution and so can be calibrated to data of any observation
length.

\item the Hamilton filter cannot fully define the entire
HMM model since the initial state probabilities are one of the key
HMM parameters in the definition (see HMM definition in section
\ref{Regime Switching Volatility Model Introduction}).

\item we cannot determine
the probability of observation sequences $p(O|M)$, since we
require the initial state probabilities. This can be understood
from the forward algorithm.

\item we cannot determine the most likely state sequence that accounts
for a given observation sequence and HMM, which can be obtained by
the Viterbi algorithm.  The Viterbi algorithm tells us the most
likely state sequence for a given observation sequence and HMM
parameters M (see Forney \cite{forneyjr1973va} for more information). 

\item without the initial state probabilities, we cannot simulate state sequences since
the initial state radically alters the state sequence and its
influence on the state sequence increases as the sequence size
decreases. Consequently we cannot validate a model's feasibility
by simulation.
\end{enumerate}
Note that  BW estimates initial state probabilities independently
of the transition probabilities, whereas in the Hamilton filter
$\eta_{i}$ is a function of estimated transition probabilities.
Hence BW is able to independently estimate more HMM parameters
than the Hamilton filter.

Thirdly, to our knowledge the Hamilton filter cannot be applied to
multivariate distributions, nor more complicated univariate
distributions than Gaussians. Particularly for financial
applications, we use multivariate data to model portfolios and
multivariate stochastic volatility is becoming an increasingly
important research area (see Bauwens et al. \cite{bauwens2006mgm}
for a survey on multivariate GARCH). Hamilton has proposed a
calibration method for univariate mixture distributions, the
Quasi-Bayesian MLE approach \cite{hamilton1991qba}, yet this
requires some prior knowledge regarding the reliability of
observations.
The GM-BW calibrates a multivariate Gaussian mixture to
multivariate data, thereby capable of modelling most empirically
observed distributions.

Fourthly, the GM-BW can calibrate time varying correlations. It is
known that correlations amongst random variables tend to be
unstable with time;  for example Buckley et al.
\cite{buckley185poa} give evidence of covariances varying with
time and model them as regime dependent. The GM-BW algorithm gives
the covariance matrix for each regime and each regime is
postulated to be linked to an economic state. Therefore, we can
model and extract information on changing correlations with
changing economic conditions. For instance, some stocks are
considered to be strongly correlated with the economic cycle
(known as cyclical stocks) e.g. British Airways, whereas other
stocks are considered independent of the economy (known as
defensive stocks) e.g. Tesco.


Finally, the BW algorithm is a complete HMM estimation method
whereas the Hamilton filter is not. Hamilton's method provides no
method or guidance as to the optimisation algorithm to apply for
finding the parameters from the non-linear filter, yet the 
solutions can be significantly influenced by the non-convex
optimisation method applied. The BW algorithm includes an
estimation method for the full HMM and a numerical optimisation
scheme. Additionally, the BW method is guaranteed to find the
local optimum.

\section{Numerical Experiment: Baum-Welch GM-HMM Calibration Results}
In this section we calibrate  a 2-state regime switching model
with 2 Gaussian components to S\&P 500 data from 1976-1996. We
fitted the model:
\begin{eqnarray}
dX/X \sim c_{j1}\mathcal{N}_{1}(u_{j1},\varphi_{j1}) +
c_{j2}\mathcal{N}_{1}(u_{j2},\varphi_{j2}), j \in \{1,2\}.
\end{eqnarray}
We set our observation to annual log returns:
\begin{eqnarray}
O_{t}=log(X(t+\Delta t)/X(t)),
\end{eqnarray}
where $\Delta t$=1 year and X(t) is the stock price.

\subsection{Procedure}
 Due to
GM-BW's  wide usage in engineering it has already been implemented
by numerous authors. We chose K. Murphy's Matlab implementation
\cite{MurphyHMMProgram} because it is considered one of the most
standard and cited GM-BW programs. It also offers many useful
features that are unavailable on other implementations e.g. the
Viterbi algorithm for obtaining state sequences.

The GM-BW algorithm finds the best GM-HMM parameters that maximise
the likelihood of the observations, however, this involves
searching a nonconvex search space and BW only finds the local
optimum. Theoretically, the globally optimal parameters can be
determined by initialising the GM-BW algorithm over every possible
starting point, then the globally optimal parameters are those
that give the highest likelihood. However, the GM-BW algorithm
finds the locally optimal
$\overline{M}=(\mathbf{\overline{A}},\mathbf{\overline{B}},\mathbf{\overline{\pi}})$
(where $\mathbf{\overline{B}}$ is parameterised by
$\overline{c}_{jk},\overline{\varphi}_{jk}$ and
$\overline{u}_{jk}$), so that the calibration problem has a
nonconvex solution search
space over thirteen dimensions. 

Due to the high dimensionality of the parameter estimation problem
for either the univariate or multivariate case, determining the
optimal parameters by initialising through different starting
points is impractical. Instead we concentrated our effort on
finding good initial parameter estimates for M. This was to
significantly increase the probability of finding the best GM-BW
solutions, particularly as initialisation strongly influences the
GM-BW optimisation \cite{liu2004eih}.
\\\\
\emph{$\mathbf{\overline{\pi}}$ Initialisation}\\
To initialise $\overline{\pi}$, we assign a probability of 0.5 to
state one if the first observed return is positive, with the
probability increasing in value the more positive it is.
\\\\
\emph{$\mathbf{\bar{A}}$ Initialisation}\\
Given the HMM structure was chosen due to its ability to capture
long term and fundamental properties, we can initialise
$\mathbf{\overline{A}}$ based on the long term and fundamental
properties we expect it to possess:
\begin{eqnarray}
\label{initialisation transition matrix} \mathbf{\bar{A}}=
\begin{pmatrix}
0.6 & 0.4   \\
0.7 & 0.3   \\
\end{pmatrix}.
\end{eqnarray}
Each state represents an economic regime so
$\mathbf{\overline{A}}$ can be interpreted as follows. The economy
in the long term follows an upward drift, so we would expect it is
more likely the HMM remains in state one rather than goto state
two, given it is already in state one -hence we assign probability
0.6.
 Similarly, if the HMM is in state two
we would expect it is far more likely to return to state one than
state two to capture the cyclical behaviour and in the long term
the economy follows an upward trend, hence we assign probabilities
0.7 and 0.3.
\\\\
\emph{GM Initialisation ($\mathbf{\overline{B}}$)}\\
The GM distribution fitting strongly affects the GM-BW algorithm
optimisation, hence it must be satisfactorily initialised for
GM-BW to provide acceptable results. However it is well known that
GM distribution fitting in general (without any regime switching)
is a non-trivial problem:
\begin{itemize}
\item there are a large set of parameters to estimate.

\item there exists the issue of uniqueness, that is for a
given non-parametric distribution there does not always
exist a unique set of GM parameter values.  

\item the flexibility of GM distributions to model
virtually any unimodal or bimodal distribution means that it
incorporates rather than rejects any noise in the data into the
distribution. Therefore GM fitting is highly sensitive to noise.

\item parameter estimation is further complicated with regime switching
and the fact  we cannot identify with certainty the (hidden) state
associated with each observation.

\end{itemize}
Rather than randomly initialise the GM parameters (as is done in
Murphy's program) we approximately divided data into each regime
by classifying positive returns as belonging to state one,
otherwise they belong to state two. We then fit a GM for each
``regime's" data using a GM fitting program used by Lund
University (stixbox).

\subsection{Results and Discussion}
We present the results of the Baum-Welch GM-HMM calibration to
S\&P 500 data from 1976-1996:
\pagebreak[4]

\begin{table}[h]
\begin{center}
\caption{Initial State Probabilities ($\pi_{i}$)}\label{Initial
State Probabilities}
\begin{tabular}{|c|c|}\hline
  State (i) & Probability \\\hline
  1 & $1 \times 10^{-6}$\\
  2 & $1-1 \times 10^{-6}$\\\hline
\end{tabular}
\end{center}
\end{table}

\begin{center}
\begin{eqnarray*}
\overline{\mathbf{A}}=
\begin{pmatrix}
  0.78 & 0.22 \\
  0.82 & 0.18\\
\end{pmatrix}
\end{eqnarray*}
\end{center}

\begin{table}[h]
\begin{center}
\caption{Mixture Means $u_{jk}$ (\%/year)}\label{Mixture Means}
\begin{tabular}{|c|c|c|}\hline
Gaussian  Component (k) & \multicolumn{2}{|c|}{State
(j)}\\\cline{2-3}
   & 1 & 2 \\\hline
 $\mathcal{N}_{1}$  & 13.0  & -4.8 \\
  $\mathcal{N}_{2}$ & 28.0  & 1.4 \\
  Overall  & 14.8  & -4.7 \\ \hline
\end{tabular}
\end{center}
\end{table}

\begin{table}[h]
\begin{center}
\caption{Mixture Standard Deviations $\sqrt{\varphi_{jk}}$
(\%/year)}\label{Volatility}
\begin{tabular}{|c|c|c|}\hline
  Gaussian Component (k) &
\multicolumn{2}{|c|}{State (j)}\\\cline{2-3}
   & 1 & 2 \\\hline
   $\mathcal{N}_{1}$ & 4.5  & 5.6 \\
  $\mathcal{N}_{2}$ & 28.0 & 110.0 \\
  Overall & 11.6  &  13.3\\\hline
\end{tabular}
\end{center}
\end{table}

\begin{table}[h]
\begin{center}
\caption{Weighting Matrix ($c_{jk}$)}\label{Weighting Matrix}
\begin{tabular}{|c|c|c|}\hline
  Gaussian  Component (k)  &
\multicolumn{2}{|c|}{State (j)}\\\cline{2-3}
  & 1 & 2 \\\hline
  $\mathcal{N}_{1}$ & 0.88 & 0.99\\
  $\mathcal{N}_{2}$ & 0.12 & 0.01 \\\hline
\end{tabular}
\end{center}
\end{table}
\newpage

From table \ref{Mixture Means} we can infer that the BW algorithm
has attributed state two as the down state since its overall mean
is negative, unlike state one. Using the Markowitz's variance
measure of risk \cite{markowitz1952ps} it is encouraging that we
can conclude that the risk level in state two is higher than in
state one (see table \ref{Volatility}), since a declining economy
(state two) is a riskier economic state. Additionally, an increase
in variance (and therefore volatility) with lower returns is
consistent with the leverage effect. The initial state
probabilities $\overline{\pi}$ strongly suggest the HMM starts in
state two and this is consistent with the data as the first year's
return (see table \ref{In Sample Regime Sequence Results}) is
relatively low (1.2\%).

The transition matrix $\mathbf{\bar{A}}$ is similar to the
transition matrix theoretically postulated (for an economy);  thus
it is consistent with our theoretical expectations. The matrix
$\mathbf{\bar{A}}$ tells us that given the model is in state one
it is likely to stay in that state most of the time, only
transitioning to state two for 22\% of the time. Additionally, in
state two the model is most likely to return to state one
(probability 0.78), thus captures the cyclical nature of the
economy.

To validate the quality of the GM-HMM calibration in terms of
state sequence generation, we ran the Viterbi algorithm. Note that
calibration under the Hamilton filter does not provide or enable
state sequence estimation. The Viterbi algorithm gave the
following state sequence result for in sample data 1976-96 in
table \ref{In Sample Regime Sequence Results}:
\newpage
\begin{table}
\begin{center}
\caption{In Sample Regime Sequence Results}\label{In Sample Regime
Sequence Results}
\begin{tabular}{|r|l|l|}\hline
Year & Regime & Empirical Annual Return (\%)\\\hline
1976 & 2 & 1.2\\
1977 & 2 & -13.4\\
1978 & 1 & 11.3\\
1979 & 1 & 13.4\\
1980 & 1 & 12.6\\
1981 & 2 & -7.3\\
1982 & 1 & 18.8\\
1983 & 1 & 11.7\\
1984 & 1 & 9.5\\
1985 & 1 & 16.5\\
1986 & 1 & 25.8\\
1987 & 2 & -6.5\\
1988 & 1 & 14.6\\
1989 & 1 & 10.1\\
1990 & 1 & 4.4\\
1991 & 1 & 17.3\\
1992 & 1 & 7.1\\
1993 & 1 & 9.3\\
1994 & 2 & -2.4\\
1995 & 1 & 30.2\\
1996 & 1 & 21.2\\\hline
\end{tabular}
\end{center}
\end{table}
The regime sequence concurs with the empirical observations;
negative or low returns were categorised into state two, whereas
other returns were classed as  state one. The results also
illustrate the behaviour of the transition matrix
$\overline{\mathbf{A}}$; generally remaining in state one (up
state) whilst occasionally entering state two, in which case it
quickly reverts back to state one. Hence the GM-HMM is able to
capture the key characteristics of the economy, namely an upward
trend and cyclicity.

The Viterbi algorithm was also applied to out of sample data from
1997-2007. Again the GM-HMM was able to satisfactorily classify
the years for each state, as shown in table \ref{Out of Sample
Regime Sequence Results}:
\newpage
\begin{table}[htbp!]
\begin{center}
\caption{Out of Sample Regime Sequence Results} \label{Out of
Sample Regime Sequence Results}
\begin{tabular}{|r|l|l|}\hline
Year & Regime & Empirical Annual Return (\%)\\\hline
1997 & 1 & 22.1\\
1998 & 1 & 26.6\\
1999 & 1 & 8.6\\
2000 & 2 & -2.1\\
2001 & 2 & -18.9\\
2002 & 1 & -27.8\\
2003 & 1 & 27.9\\
2004 & 1 & 4.3\\
2005 & 1 & 8.0\\
2006 & 1 & 11.6\\
2007 & 2 & -2.6\\\hline
\end{tabular}
\end{center}
\end{table}

\section{Conclusions}
This paper has shown the advantages of Baum-Welch calibration over
the  standard Hamilton filter method. Not only does the Baum-Welch
method offer a complete calibration procedure but also is able to
estimate the full set of HMM parameters, unlike the Hamilton
filter. We have also validated the usage of the Baum-Welch method
through numerical experiments on S\&P 500 data in and out of
sample.

\newpage
\bibliographystyle{plain}
\addcontentsline{toc}{section}{References}
\bibliography{Ref}

\end{document}